# Corrections to the Saffman-Delbrück mobility for membrane bound proteins

Ali Naji*, Alex J. Levine [†], and P.A. Pincus [‡]

*Department of Physics and Department of Chemistry & Biochemistry, University of California, Santa Barbara CA 93106; [†]Department of Chemistry & Biochemistry and California Nanosystems Institute, University of California, Los Angeles, CA 90095; [‡]Marterials Research Laboratory, University of California, Santa Barbara, CA 93106; [‡]Departments of Physics and Materials, Biomolecular Science & Engineering, University of California, Santa Barbara, CA 93106. Email: alevine@chem.ucla.edu

ABSTRACT  Recent experiments by Y. Gambin et al. [PNAS **103**, 2098 (2006)] have called into question the applicability of the Saffman-Delbrück diffusivity for proteins embedded in the lipid bilayers. We present a simple argument to account for this observation that should be generically valid for a large class of transmembrane and membrane bound proteins. Whenever the protein-lipid interactions locally deform the membrane, that deformation generates new hydrodynamic stresses on the protein-membrane complex leading to a suppression of its mobility. We show that this suppression depends on the protein size in a manner consistent with the work of Y. Gambin et al.

The diffusivity of transmembrane proteins is a fundamental biophysical parameter controlling the dynamics of protein-protein interactions in the cell membrane. These dynamics underlie such processes as endocytosis and signal transduction (1). Understanding the size dependence of the diffusivity of membrane bound proteins is rather subtle. Saffman and Delbrück (SD) (2) originally demonstrated the significant differences between lateral diffusion in membranes and the better understood problem of diffusion in a bulk solvent. In the membrane, the diffusion constants are only weakly dependent on the size of the diffusing particle, while in bulk solvent the diffusion constant depends inversely on particle size. Although some data appear to support the Saffman-Delbrück theory (3,4,5), more recent experiments exploring the diffusivity of transmembrane proteins over a larger size range (6) using *in vitro* lipid bilayers show a much stronger protein-size dependence than is consistent with our current understanding of membrane hydrodynamics (2,7,8). These data suggest that the diffusivities of the proteins depend inversely on their size for a variety of proteins and protein aggregates covering about one decade of inclusion radius, and are clearly inconsistent with the SD result.

In this Letter, we address this puzzling discrepancy between theory and experiment by proposing that accounting for local membrane deformations caused by embedded proteins can resolve this conflict. We reexamine the mobility µ of a protein in the lipid bilayer. The mobility defines the linear relationship between a particle's velocity $\vec{v}$ and the force $\vec{F}$ applied to it via the relation

$$\vec{v} = \mu \vec{F}.$$

From the protein mobility and the Stokes-Einstein relation, $D = \mu k_B T$, one determines its diffusion constant in the membrane.

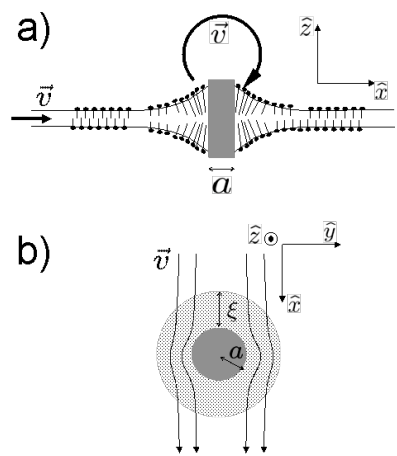

Figure 1: a) Side-view of the membrane with an embedded protein (grey), which creates a bump generating additional flows in the solvent (black arrows) leading to extra dissipation; b) Top-view of the membrane showing the protein (grey) and the perturbed membrane surrounding it (hatched).

It is well-known that the mobility of a rigid, spherical particle of radius *a* in a three-dimensional solvent having viscosity $\eta$ is given by the Stokes result, $\mu = 1/(6\pi\eta a)$, which has an inverse dependence on the particle radius. The mobility of the same particle when embedded in a fluid membrane however is more complex. There the particle moves through an effectively two-dimensional liquid that is coupled to the surrounding three-dimensional solvent by the requirement that there be no slip at the interfaces between the lipid membrane and the aqueous solvent. The hydrodynamic coupling between flows in the effectively two-dimensional fluid and the surrounding solvent introduces an inherent length scale into membrane hydrodynamics – the SD (2) length $\ell = \eta_m/\eta$, which is set by the ratio of the two-dimensional membrane viscosity $\eta_m$ to that of the surrounding bulk solvent $\eta$. In contrast,

Published in Biophysical Journal (Biophysical Letters) **93**, L49 (2007)

the usual low-Reynolds-number hydrodynamics in a bulk liquid is a scale-free theory.

The introduction of this extra length scale profoundly modifies the mobility of particles embedded in the membrane. Because of this, the mobility of a particle of radius $a$ in the membrane is given by

$$\mu_{membrane} = \frac{1}{4\pi\eta_m} f(\ell/a) ,$$

where the scaling function $f$ has the following limits for large (2) and small (7) arguments:

$$f(x) = \begin{cases} \pi x/4, & x \ll 1 \\ \ln(x) - \gamma_E, & x \gg 1 \end{cases}$$

and $\gamma_E$ is the Euler-Mascheroni constant. While the mobility of a sphere in a bulk fluid depends inversely on its radius, its mobility in a fluid membrane has only a weak, logarithmic dependence on particle size for particles much smaller than $\ell$. For proteins, the applicable limit is $\ell/a \gg 1$, which suggests that, to a good approximation, all membrane bound proteins and even the constituent lipids of the membrane should have essentially the same diffusion constant.

Recent experiments on the diffusion of lipid domains on giant unilamellar vesicles quantitatively support this form of the lateral mobility of embedded objects in membranes (5), but analogous experiments on membrane-bound protein mobilities do not (6). Taken together these data suggest a resolution of this conflict. The mobilities computed above depend on only a few simple assumptions regarding mass and total momentum conservation: thus they appear unassailable. It is well known, however, that membrane bound proteins typically perturb the membrane structure locally (9-12). This local perturbation may take many forms including local changes in membrane height or thickness involving local oligomeric chain stretching, local membrane curvature, tilt of the lipids, or changes in local lipid composition relative to that the far field (for mixed lipid systems). Below we show that these membrane perturbations that must be transported along with the proteins can shift the mobility of these composite objects from the SD form to one consistent with $D \sim 1/a$ scaling. This effect arises from either the enhanced dissipation associated with modifications of the flows in the bulk solvent caused by the protein-induced height or bending deformations, or the enhanced dissipation occurring within the membrane itself in cases where the protein generates local changes in composition, chain stretch, or tilt order. These two scenarios are not mutually exclusive and both give the same $D \sim 1/a$ scaling, but, as they rely on somewhat different reasoning, we present the arguments independently.

*Height mismatch and bulk hydrodynamics* – If the protein has a hydrophobic mismatch with the membrane thickness, it will generate a bump on the surface having a lateral dimension on the order of the radius of the protein or the membrane thickness $h$ (10-12). We now estimate the effective mobility $\mu_{eff}$ of the protein and associated membrane deformation (bump) by considering the power dissipated, $P = Fv$, when this complex is moved at constant speed $v$ in the membrane in response to an applied force $F$. Using the definition of mobility, the power input required to move the protein is

$$P_{in} = v^2/\mu_{eff} .$$

In steady state, this must equal the sum of the power dissipated in the membrane and any additional viscous dissipation arising from the perturbation of the velocity field in the bulk solvent by the bump. This latter is given by

$$\int \eta (\nabla v)^2 d^3x \sim C\eta v^2 a ,$$

where the volume integral of the product of the velocity gradient and viscous stress determines the extra power dissipated in the fluid due to the perturbation of the bulk velocity field around the deformed membrane (13) and $C$ is a constant of order unity. To estimate the above integral, we recognize that perturbed velocity field extends a characteristic distance of order the protein radius $a$. More precisely, it is the mismatch between the size of the hydrophobic protein domain and the membrane height that drives the added dissipation. We assume that this mismatch scales with the protein dimension. See Fig. 1a. If we assume that the effect of the membrane deformation on the internal membrane flows is minimal, we may add this power dissipation to that associated with dissipation in the flat membrane to write

$$P_{out} = v^2/\mu_{membrane} + C\eta v^2 a .$$

Equating the power input and output, we find that

$$\mu_{eff} = \frac{\mu_{membrane}}{1 + C\eta a \mu_{membrane}} ,$$

which is the desired particle-radius dependence when $C\eta a\mu_{membrane} \gg 1$.

*Dissipative protein-lipid interactions and membrane hydrodynamics* – It is also possible that the principal additional dissipative stresses are associated with degrees of freedom internal to the membrane. Such dissipation may be related to chain stretching or tilt near the protein, or to local demixing of the constituent lipid species resulting from differential affinities between the protein and the various lipid species.

We now estimate the power dissipated in the membrane. As shown in Fig 1b, we posit that the disruption of membrane structure occurs within a distance $\xi$ of the protein. Working in the reference frame of the protein, lipids flow into this modified zone and undergo some entropy-generating (i.e. dissipative) conformational change in some boundary layer around the zone of width $\delta\xi$ where the power dissipated per lipid is $p_{lipid} = fv = \varepsilon'' v^2$. Since the dissipative forces $f$ must be odd under time reversal they must be linear in the rate of lipid deformation, which is linear in velocity. Using the area density of the lipids $\rho$ and the area of affected lipids, $2\pi(a+\xi)\delta\xi$, to determine the number of lipids involved in the extra power dissipation we find

$$P_{out} = v^2/\mu_{membrane} + \rho\, 2\pi(a+\xi)\delta\xi\varepsilon''v^2.$$

If we now assume that the zone of lipid deformation is small compared to the radius of the protein and that the boundary layer of this zone is comparable to the width of the zone itself, $\xi \approx \delta\xi << a$, we may simplify the above expression as $P_{out} = v^2/\mu_{membrane} + v^2\Gamma a$, where $\Gamma = 2\pi\rho\varepsilon''\xi$. Equating the power dissipated to the power input as before, we arrive at another expression for the effective mobility of the protein membrane complex

$$\mu_{eff} = \frac{\mu_{membrane}}{1+\Gamma a\,\mu_{membrane}}.$$

In this latter case we cannot estimate the magnitude of $\Gamma$ and thus cannot make predictions for the protein size at which one should expect to see the inverse $a$ scaling of the protein's diffusion constant. In fact, this cross-over size will most likely depend on protein-specific details of the protein-lipid interactions and the lipid composition of the membrane. In the former case, where the extra dissipation occurs entirely in the bulk solvent, we can make quantitative estimates of the effect. Examining the effective mobility predicted in this case, we see that the inverse $a$ scaling should occur where $\ell/a < (2+C/12\pi)$. In the case of very viscous membranes where $\ell$ is much larger than protein radius, we cannot expect that dissipation in the less viscous bulk solvent to dominate the total dissipated power as the protein moves through the membrane. In the experiments of Gambin et al. (6), however, typically $\ell \approx 100a$ so we expect dissipative protein-lipid interactions to account for the size dependence of the diffusivities.

We have shown that one may account for the experimentally observed failure of the Saffman-Delbrück diffusivity of membrane bound proteins by positing that the protein carries with it a locally deformed patch of membrane. This local deformation will generate extra flows in the bulk solvent if the protein creates a bump or depression in the membrane; if the protein modifies the internal structure of the lipids in its immediate vicinity, then there is enhanced dissipation in the membrane as the deformation is dragged by the protein. As long as the power dissipated in the membrane or in the surrounding solvent arising from this membrane deformation is at least comparable to the dissipation in the usual flows of the unperturbed membrane, one will observe an inverse radius dependence of the protein diffusivity.

Recent simulations of inclusion mobility (14) have also found deviations from the SD mobility of inclusions. There it was found that $\mu \sim 1/a^2$ because of the dissipation enhancement coming from internal soft modes of the inclusion. That work shows yet another way in which extra internal degrees of freedom shift the mobility of the object. These inclusions did not deform the membrane in ways that we suggest here; new simulations having these effects are clearly desirable. Experiments on lipid domain mobilities find agreement with the Saffman-Delbrück expression (5). There one should not expect large perturbations of the surrounding membrane by these domains. Transmembrane proteins, on the other hand, are known to generate static deformations of the surrounding membrane. Little work has been done on examining the dynamic effects of such membrane perturbations. Our simple heuristic analysis suggests that both futher theoretical work and more local examinations of protein dynamics in membrane are required to better understand protein transport properties in lipid bilayers.

This study was partially carried out at the Aspen Center for Physics and we would like to express our appreciation for their hospitality. P.A.P. acknowledges support from grants NSF DMR-0503347 and DMR-0710521. We also wish to acknowledge important discussions with N. Taulier, F. Brown, W. Urbach, R. Netz, and J. Hardin.